# Optical Diffraction Tomography based on 3D Physics-Inspired Neural Network (PINN)

Ahmed B. Ayoub[1,*], Amirhossein Saba[1], Carlo Gigli[1], Demetri Psaltis[1]

[1] Ecole Polytechnique Fédérale de Lausanne, Optics Laboratory, CH-1015 Lausanne, Switzerland.
* Email: ahmed.ayoub@epfl.ch

Abstract: Optical diffraction tomography (ODT) is an emerging 3D imaging technique that is used for the 3D reconstruction of the refractive index (RI) for semi-transparent samples. Various inverse models have been proposed to reconstruct the 3D RI based on the holographic detection of different samples such as the Born and the Rytov approximations. However, such approximations usually suffer from the so-called "missing-cone" problem that results in an elongation of the final reconstruction along the optical axis. Different iterative schemes have been proposed to solve the missing cone problem relying on physical forward models and an error function that aims at filling in the k-space and thus eliminating the "missing-cone" problem and reaching better reconstruction accuracy. In this paper, we propose a different approach where a 3D neural network (NN) is employed. The NN is trained with a cost function derived from a physical model based on the physics of optical wave propagation. The 3D NN starts with an initial guess for the 3D RI reconstruction (i.e. Born, or Rytov) and aims at reconstructing better 3D reconstruction based on an error function. With this technique, the NN can be trained without any examples of the relation between the ill-posed reconstruction (Born or Rytov) and the ground truth (true shape).

## Introduction

Optical diffraction tomography (ODT) is a label-free imaging technique that can be used to characterize different samples by quantifying their 3D refractive index (RI) [1-10]. Unlike fluorescent-based imaging, ODT does not suffer from photo-bleaching, photo-toxicity, and thus can permit for imaging over a long span of time. As it does not require staining or fixation of cells, ODT can be used for many applications that involve live imaging of samples such as the red blood cells membrane dynamics [7], biological tissues [8], and c-elegans [9].

To reconstruct the 3D RI maps of such samples, different holographic images are obtained by illuminating the sample from different illumination angles. The complex field is then retrieved from those holograms to reconstruct the 3D RI map by means of inverse models including Born, and Rytov approximations [10]. The main limitation in such reconstructions is the infamous missing-cone problem which arises from the limited numerical aperture (NA) of the objective lens in the imaging setup. This results in two associated consequences: (1) under-estimation of the RI values, and (2) elongation of the 3D RI reconstruction along the optical axis [11, 12].

Several approaches have been proposed to solve the missing cone problem including the use of iterative schemes that enhance the 3D RI accuracy by using a physical forward model that simulate optical beam propagation through the current estimate of the 3D RI and the experimentally measured 2D complex field [12-14]. By updating the current estimate in order to minimize the error between the field from the forward model and the experimentally measured field, the 3D RI map is obtained. Other approaches were concerned with intensity-only measurement where the experimental measurements do not involve the use of a separate reference arm as in conventional microscopy. In these works, the error function is computed between the intensity that is experimentally measured and the one calculated with the forward model as proposed in [15-16].

A new paradigm to ODT was proposed in reference [17] using a Deep neural network (DNN) to solve the missing cone problem. The DNN was trained with hundreds of examples for red blood cell digital phantoms where the input is the 3D RI map using the Rytov approximation while the output of the network is the 3D actual RI of the red blood cell. The training process includes simulated phantoms while the testing was based on real measurements. Other recent works that involve solving the missing cone problem were proposed in [18-20] where the neural networks are usually trained on synthetic data and then tested on real samples to validate its accuracy.

Physics-inspired neural networks (PINN) differs than the previously mentioned NN in that they do not need examples of input-output pairs to learn. Instead, a physical model is used to control the training process and thus avoid the need to collect a large number of examples of the ground truth and computational memory for to store the large database. PINN has been used in different fields including heat transfer [21], solving partial differential equations (PDE) [22], solving-Maxwell's equations [23], and phase imaging [24]. In this paper, we propose a new approach for ODT by utilizing PINN for 3D reconstruction of objects. Using this technique only one ill-posed 3D reconstruction we could solve the missing-cone problem. A schematic diagram depicting the approach is shown in Fig. 1.

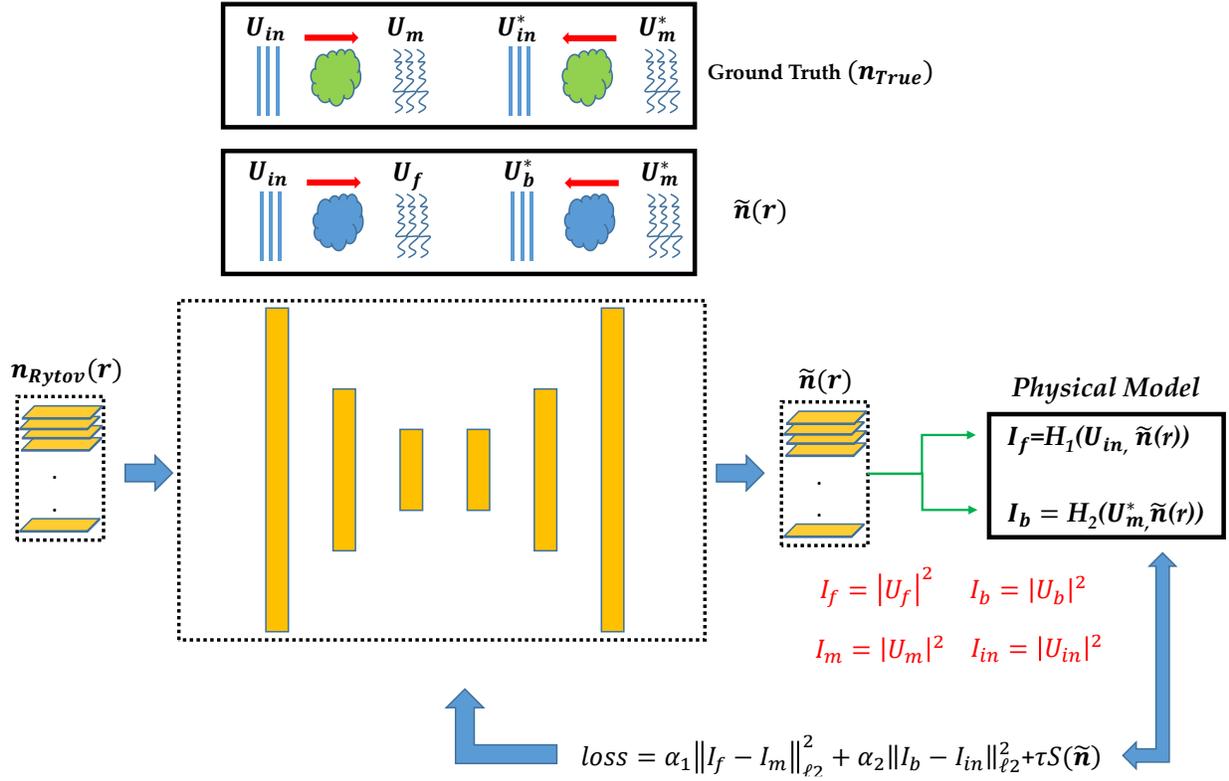

Figure 1. Overall Idea. The input to the neural network is the 3D RI Rytov reconstruction and then the output of the NN is the current guess of the RI $\tilde{n}(r)$ which is then used in the BPM model for both forward and backward propagation using the actual incident and actual total field. A loss function is calculated for both the forward and backward propagation and this loss function is back-propagated through the NN to update the weights of the NN resulting in a better reconstruction quality.

# Methods

## Network Architecture

The DNN architecture is shown in Fig. 2. The DNN is a U-Net [25] with the encoder and the decoder arms each composed of 5 layers. The input and the output of the network is a 3D volume of size 64x64x64 that contains the 3D RI map. The input to the network is the Rytov approximation [10] while the output keeps updated every epoch based on a loss function that is decided by the physical model as will be explained later. 3D convolutions are applied twice at each layer of the U-Net with kernel size of 3x3x3 followed by 3D max-pooling of 2x2x2. Zero padding is added to keep the size unchanged after each convolution. Concatenation along the last axis (i.e. channels) is added to localize important features in the 3D map along the U-Net. For the optimization, Adam optimizer was used with a decaying learning rate (additional details in the results section). ReLU was used as an activation function. The U-Net architecture was implemented using Tensorflow and was run on desktop computer (Intel Core i7-7700K CPU, 4.2 GHz, 32 GB RAM) with a graphics processing unit (NVIDIA GeForce GTX 1080 GPU).

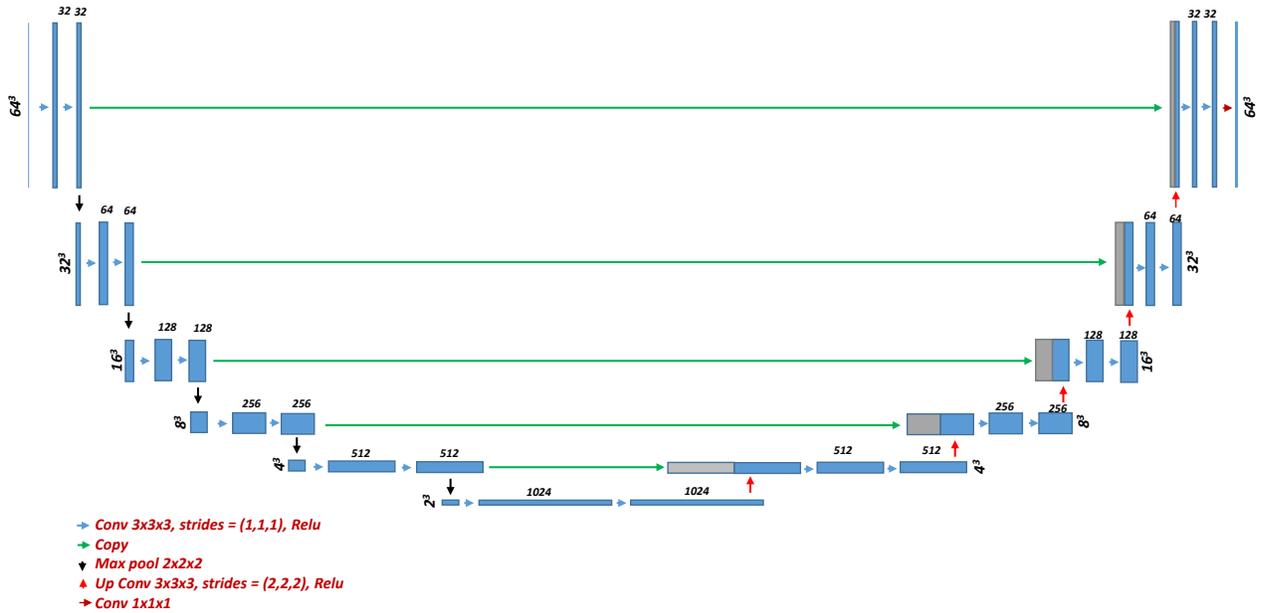

Figure 2. U-Net architecture.

## Physical Model

The physics of phase conjugation states that the distortion introduced to an incident wave due to propagation through an inhomogeneous, lossless medium is undone by conjugating the distorted wave and then launching it backwards through the same medium. In this way, the incident wave is reproduced. We use this property to derive an error function for the NN that generates the 3D estimate of the object. The transmitted wave is experimentally measured holographically and digitized. The digitized wavefront is then phase conjugated and digitally propagated backwards through the current estimate of the 3D index distribution of the sample. If the current estimate is a faithful reproduction of the ground truth, then the light distribution of the conjugate wave is a faithful reproduction of the incident wave. We obtain the error function by calculating the difference in the intensities of the actual incident field and the digital reconstruction. This error signal is then used to train the NN via error back propagation. We can also derive a similar error function by comparing the intensity of the experimentally measured transmitted field with a digital calculation for the field transmitted through the current estimate (Figure 3). We found that we get enhanced performance when using both error functions simultaneously.

The physical model is based on beam propagation model (BPM) to simulate the light propagation across the current guess of the 3D RI map. Using BPM, the scattering medium can be refined into fine steps where at each step the optical wave propagation can be divided into two consecutive processes: propagation step and phase modulation step as described in the following equation:

$$A(x,y,z+dz) = F_{2D}^{-1}\{F_{2D}[A(x,y,z)] \cdot e^{-\frac{jdz[k_x^2+k_y^2]}{k+k_z}}\} e^{jk_0 \Delta n(x,y,z)[dz/\cos(\theta)]} \tag{1}$$

where $A(x,y,z)$ is the slowly-varying envelope of the field $U(x,y,z) = A(x,y,z)e^{jkz}$. $F_{2D}$, $F_{2D}^{-1}$ are the Fourier transform and the inverse Fourier transform respectively. $\Delta n$ is the refractive index contrast between the sample and the surrounding medium ($n_0$). $k_z = \sqrt{k^2 - k_x^2 - k_y^2}$, $k = k_0 n_0$, $k_0 = \frac{2\pi}{\lambda}$, $\lambda$ is the wavelength, and $\theta$ is the illumination angle for the respective field given by $\theta = cos^{-1}(\frac{k_{in}^z}{k})$, where $k_{in} = (k_{in}^x, k_{in}^y, k_{in}^z)$ is the incident k-vectors.

As shown in Fig. 3(a), the retrieved field $U_f$ is a function of the input illumination field and the current estimate of the 3D refractive index from the neural network where $H_1$ represents the optical wave propagation in equation (1) until we reach the end of the propagation window $z = L$, where $L$ is the size of the propagation window.

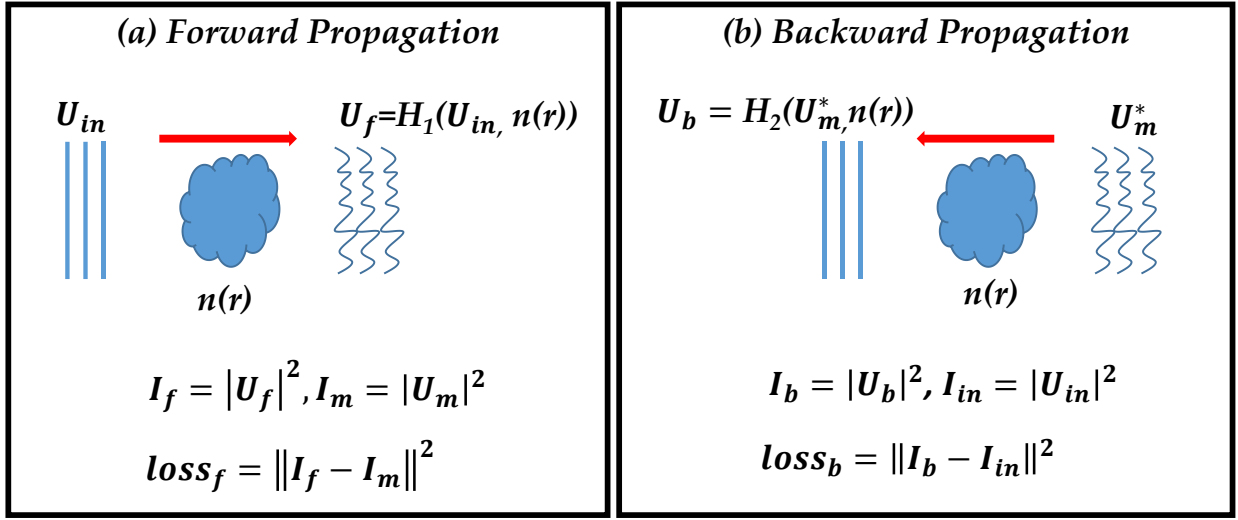

Figure 3. Physical Model. (a) Forward propagation, (b) backward propagation.

In Fig. 3(b), the phase conjugation and optical wave propagation can be explained as follows:

$$U_m = T_{io} U_{in} \tag{2}$$

where $T_{io}$ represents the transmission matrix describing the optical propagation from input to output through the physical sample.

By taking the complex conjugate of the output field and back-propagating through the current estimate of the 3D RI reconstruction $T'_{oi}$, we get:

$$U_b^* = T'_{oi} U_m^* = T'_{oi}(T_{io}U_{in})^* = T'_{oi} T_{io}^* U_{in}^* \tag{3}$$

From equation 3 we conclude that if the current estimate of the 3D RI is identical to the actual RI map (i.e. $T'_{oi} = T_{oi}$, where $T_{oi}$ is the transpose of the transmission matrix $T_{io}$), and assuming a perfect time-symmetric system ($T_{io}^* = T_{oi}^{-1}$) then we would retrieve the complex conjugate of the incident field ($U_b^* = T_{oi}T_{io}^*U_{in}^* = T_{oi}T_{oi}^{-1}U_{in}^* = U_{in}^*$). Otherwise, there would be a mismatch between the retrieved field and the incident field.

The loss function for the forward propagation is then given as:

$$loss_f = \left\| I_f - I_m \right\|_{\ell 2}^2 \tag{4}$$

where $I_f = |U_f|^2$, and $I_m = |U_m|^2$.

Similarly,

$$loss_b = \| I_b - I_{in} \|_{\ell 2}^2 \tag{5}$$

where $I_b = |U_b|^2$, and $I_{in} = |U_{in}|^2$.

The total loss function is then given by:

$$loss = \alpha_1 loss_f + \alpha_1 loss_f + \tau TV(f) \tag{6}$$

where $\alpha_1, \alpha_2$ are constants, $TV(f) = |\nabla_x f| + |\nabla_y f| + |\nabla_z f|$ and $f$ represents the RI contrast of the sample. $\nabla_x, \nabla_y, \nabla_z$ are the partial derivatives along $x, y, z$, respectively.

**Simulation setup**

In this digital experiment. We used Lippmann-Schwinger (LS) as a forward model to generate different projections as ground truth for each digital phantom [14, 26]. For this 100 different illumination angles were assumed in a circular pattern. LS was used for 2 reasons: (1) It is highly accurate, (2) to remove any bias during the reconstruction process if the data was generated using similar model to BPM. The data was generated on a desktop computer (Intel Core i7-7700K CPU, 4.2 GHz, 32 GB RAM) with a graphics processing unit (NVIDIA GeForce GTX 1080 GPU). The LS equation in the integral form is expressed as:

$$U_t = U_{in} + \int G(r' - r) F(r') U_t(r') dr' \tag{7}$$

, where $U_t$ is the total field, $U_{in}$ is the incident field, $G(r)$ is the green's function, and $F(r)$ is the scattering potential given as: $F(r) = \frac{k_0^2}{4\pi^2}(n^2(r) - n_0^2)$.

The numerical propagation is divided into 2 steps:

$$U_t = (I - GF)^{-1} U_{in} \tag{8}$$

$$E_{meas} = \tilde{G} F U_t + U_i \tag{9}$$

where $I$ is the identify matrix, $\tilde{G}$ is the green's function that calculates the field $E_{meas}$ at the sensor position. $U_t$ and $U_{in}$ are the total and incident field discretized in the region of interest. In Eq. (8), we compute the discrete total field $U_t$ in the region of interest by inverting a matrix using BiConjugate Gradients Stabilized Method that iteratively compute the inverse of the matrix $(I - GF)$ [27]. We then compute the 2D field at the sensor position using $\tilde{G}$ and $U_i$ which the incident 2D field. LS model is considered a highly accurate model since no further approximations are accounted for beyond the scalar approximation. In all the simulation, wavelength of 532 nm is assumed, $n_0 = 1.334$, pixel size = 80 nm.

## Results

At each epoch, the weights of the neural network is updated by going from the Rytov to a current estimate of the index which is then inserted into the BPM model. By using the total field and the incident field to calculate the loss functions, we can propagate this error back through the neural network where the weights are being updated. By doing so at every epoch, the weights keeps updated until the error converges. Fig. 4 shows the refractive index map at different iterations for a micro-sphere with diameter around 2 μm and index contrast of 0.1 immersed on water ($n_0 = 1.334$). The input volume has size of 64x64x64 pixels.

As we can observe, at the first few iterations (i.e. 30 iteration) we see that the reconstruction suffers from the missing cone problem due to the limited illumination angle range. However, with the help of BPM as the physical model, we can calculate the loss function between the field from BPM and the actual field from LS in our case to update the weights and thus iteratively solve the missing cone problem where the elongation along the optical axis (i.e. YZ, and XZ) is significantly suppressed at 3000, and 9000 iterations as compared to the first few iterations as shown at 30 iterations. We can verify this as well by the decreasing loss function as a function of the number of iterations which is a sign of convergence.

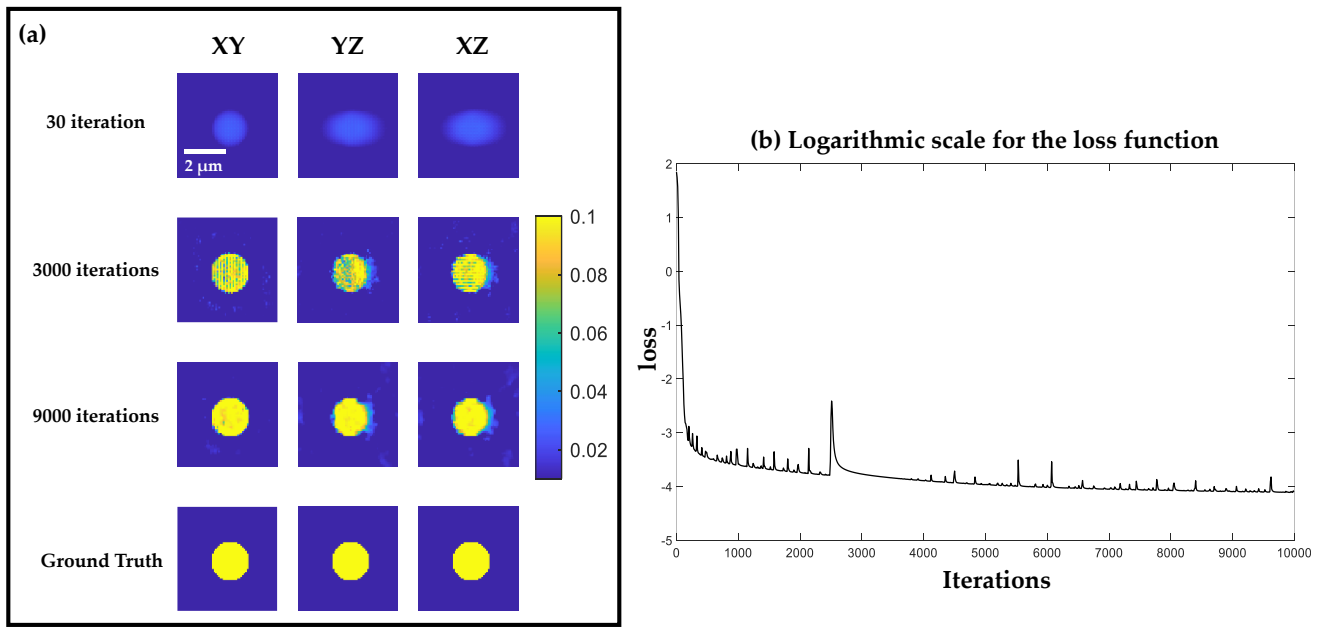

Figure 4. 3D RI Reconstructions for the microsphere. (a) XY, YZ, and XZ cross-sections of the 3D RI map at different iterations while the last row represents the ground truth. (b) The log scale of the loss as a function of the iterations where we clearly see the convergence of the loss as we progress in the iteration number.

To assess the proposed technique further, we simulated a digital phantom that has more features with varying refractive index contrast as shown in Fig. 5. As observed earlier, the loss is converging and 3D RI output from the U-Net is getting closer to the ground truth as the number of iterations keeps progressing.

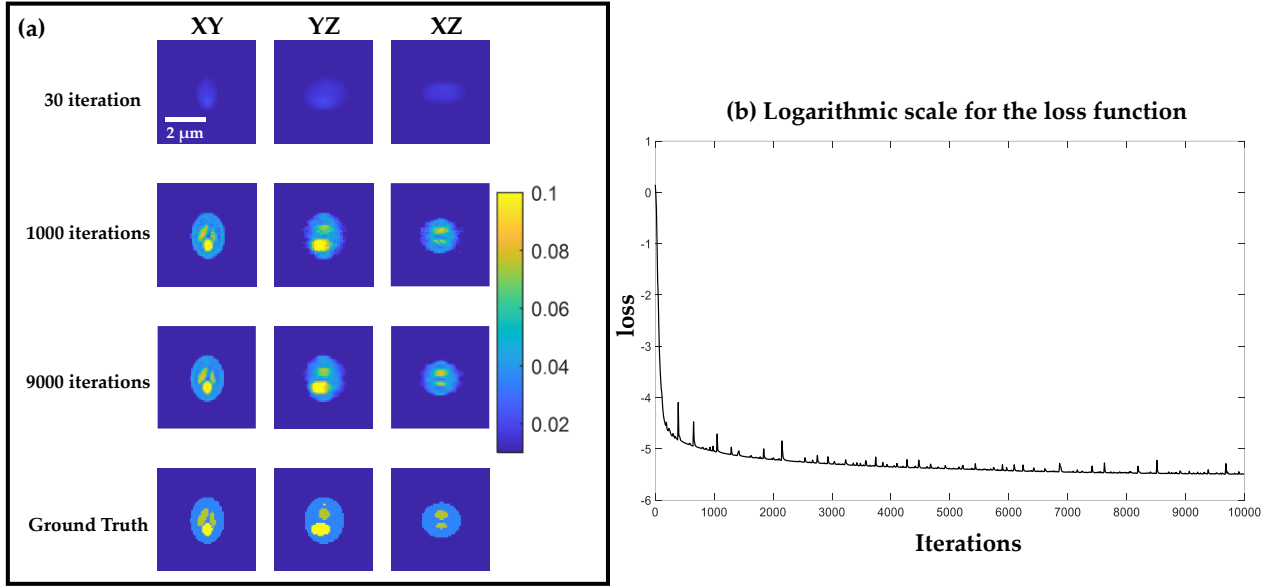

Figure 5. 3D RI Reconstructions for the digital phantom. (a) XY, YZ, and XZ cross-sections of the 3D RI map at different iterations while the last row represents the ground truth. (b) The log scale of the loss as a function of the iterations where we clearly see the convergence of the loss as we progress in the iteration number.

Table 1 shows the parameters used to reconstruct the 3D RI for the two objects.

Table 1. Optimization parameters for the microsphere and the digital phantom.

| Sample | $\alpha_1$ | $\alpha_2$ | $\tau$ | Learning rate (LR) |
|---|---|---|---|---|
| Microsphere | 0.9 | 0.1 | 1.2 | $2*10^{-5}$ |
| Phantom | 0.9 | 0.1 | 1.2 | $2*10^{-5}$ |

## Conclusion

In this work, we presented a physics-inspired 3D neural network approach to solve the missing cone problem in ODT. Instead of using thousands of examples to learn, we instead used a physical model based on BPM to supervise the neural network. By starting from the Rytov reconstruction, we showed how the U-Net is able to learn with the guidance of the physical model.

## Funding